# Interplay of Cooper and Kondo singlet formations in high-$T_c$ cuprates

Tsuyoshi Sekitani[a], Michio Naito[b], Hisashi Sato[b], and Noboru Miura[a,*]

[a] Institute for Solid States Physics, University of Tokyo, Kashiwanoha, Kashiwa-shi, Chiba, 277-8581, Japan

[b] NTT Basic Research Laboratories, Morinosato-Wakamiya, Atsugi-shi, Kanagawa, 243-0198, Japan

**abstract**

We performed high-field magnetotransport experiments up to 50 T on strained $La_{1.85}Sr_{0.15}CuO_4$ films. Two films with different strain, one expansively strained with $T_c \sim 26$ K and the other compressively strained with $T_c \sim 44$ K, show a striking contrast in the normal-state transport: the former has a prominent low-temperature upturn while the latter has not. Identifying the low-temperature upturn as due to the Kondo effect, our experimental observations indicate that Cooper and Kondo singlet formations are competing in high-$T_c$ cuprates: Kondo singlet formation hinders Cooper pair formation. Based on these results, we propose a new scenario for the electronic phase diagram of $La_{2-x}Sr_xCuO_4$. According to our scenario, underdoped insulating $La_{2-x}Sr_xCuO_4$ can be regarded as a "Kondo insulator".

**Contact address**

Noboru Miura

5-1-5, Kashiwanoha, Kashiwa-shi, Chiba 277-8581, Japan.
Tel: +81-471-36-3335, Fax: +81-471-36-3335
E-mail: miura@issp.u-tokyo.ac.jp

**Introduction**

In the 20th century, two significant many-body phenomena due to spin singlet formation were discovered in the field of solid state physics: superconductivity and the Kondo effect. Superconductivity arises from Cooper pair (singlet) formation between conduction electrons, which leads to perfect conductivity below a critical temperature ($T_c$). The Kondo effect arises from Kondo singlet formation between a conduction electron and a local spin, which leads to suppression of the conductivity (localization) below a certain temperature ($T_K$: this is not a phase-transition but a cross-over temperature). The interplay between the Kondo effect and superconductivity has been the subject of numerous theoretical and experimental studies since the 1960's [1]. In this paper, we report our recent magnetotransport experiments on $La_{2-x}Sr_xCuO_4$ (LSCO) thin films, which clearly indicate that that Kondo effect and superconductivity are competing in this compound.

High-field magnetotransport is one of the most powerful probes for investigating Kondo states. Since the magnetic field acts so as to align spins and thereby to dissociate Kondo singlets, the interplay between the Kondo interaction and the Zeeman interaction provides valuable insights into the Kondo states. High magnetic fields exceeding the upper critical fields are also required to unveil the "normal" ground state of high-$T_c$ superconductors. In the present study, high-field magnetotransport measurements up to 50 T, using a pulse magnet energized by a capacitor bank of 900 kJ (5-10 kV), were performed on $La_{1.85}Sr_{0.15}CuO_4$ films. LSCO films were prepared by molecular beam epitaxy, whose details are described in ref.[2]. $La_{2-x}Sr_xCuO_4$, a prototype high-$T_c$ superconductor, has an optimum $T_c$ of ~ 37 K for bulk specimens under ambient pressure [3]. For films, however, $T_c$ can be varied by epitaxial strain, as has been demonstrated recently by our group and other groups [4-6]. For high-field experiments, we prepared two $La_{1.85}Sr_{0.15}CuO_4$ films with identical nominal chemical composition but with different epitaxial strain. One ("high-$T_c$") film, on a $LaSrAlO_4$(001) substrate, was compressively strained in plane (*a*-axis) and expansively strained out of plane (*c*-axis) by the Poisson effect, and has $T_c$ ~ 44 K. The other ("low- $T_c$") film, on $SrTiO_3$(001), was strained in the opposite way, and has $T_c$ ~ 26 K. The lattice constants of these films are listed in Table 1.

The two LSCO films show quite different magnetotransport properties. Fig. 1 shows the temperature dependences of resistivity in magnetic fields applied parallel to the *c*-axis. With increasing magnetic field, the superconductivity is gradually suppressed, and the low-temperature normal behavior is unveiled accordingly. The normal-state resistivity of the "high-$T_c$" film is rather normal and metallic down to the lowest temperature. In contrast, the normal-state resistivity of the "low-$T_c$" film shows an anomalous semiconducting upturn (d$\rho$/d$T$<0) with a resistivity minimum at around 30 K. Low-temperature semiconducting upturns with a resistivity minimum have frequently been observed in high-$T_c$ cuprates [7, 8, 9]. As the origin of this upturn, our recent systematic magnetotransport studies on the electron-doped superconductors (Nd,Ce)$_2$CuO$_4$, (Pr,Ce)$_2$CuO$_4$, and (La,Ce)$_2$CuO$_4$, where a similar anomalous upturn was observed at low temperatures [10, 11], have provided strong evidences pointing to Kondo scattering due to Cu$^{2+}$ local spins. The evidences are (1) the log $T$ dependence of the upturn (localization effect due to singlet formation), (2) the saturation of the upturn at lower temperatures (unitarity limit of scattering), and (3) the suppression of the upturn by high magnetic fields (delocalization effect due to singlet dissociation by magnetic field). The normal-state resistivity of the "low-$T_c$" LSCO film also follows log $T$ behavior and saturates at lower temperatures as shown in Fig. 2. However, the suppression of the upturn by high magnetic fields (negative magnetoresistance) has not been observed, most likely because the magnetic field of 50 T is not sufficient to dissociate the Kondo singlets in LSCO, which seem to be more strongly bounded than in electron-doped superconductors. Experiments with higher magnetic fields of up to ~ 100 T, which we are planning, may clarify this issue.

Here we give a rough estimate for the Kondo temperature ($T_K$) and the Kondo magnetic field ($H_K$). Following the procedure by Samwer and Winzer [13] to define $T_K$ as the temperature at the half-height of the Kondo resistivity step, the $T_K$ of "low-$T_c$" LSCO is ~ 20 K. Then the scaling relation, $k_B T_K = S g_{eff} \mu_B H_K$, gives $H_K$ ~ 30 T, assuming $S = 1/2$ and $g_{eff} = 2$. However, the actual value for $H_K$ of "low-$T_c$" LSCO seems to be much larger than ~ 30 T. This discrepancy may partly be explained by supposing $g_{eff} < 2$. However, it may be better to postpone quantitative discussions to later publications since a single-impurity-scattering approximation apparently does not apply to the present case.

Given the Kondo scenario as the origin for the low-temperature upturn, our experimental observation indicates that the Kondo effect prevails in the "low-$T_c$" film, while it is practically negligible in the "high-$T_c$" film. This seems to indicate that *stronger Kondo interaction lowers superconducting transition temperature $T_c$*, or in other words, *Kondo singlet formation hinders Cooper pair formation*, which is the main argument of this article. The interplay between the Kondo effect and superconductivity for low-$T_c$ superconductors has been extensively investigated theoretically and also experimentally. However, it has not yet been considered seriously for high-$T_c$ cuprates. Both the Kondo interaction $H_K$ and the superconducting pairing interaction $H_{pair}$ are involved in this problem. In the case that $H_K$ dominates $H_{pair}$, conduction electrons pair up with local spins, resulting in Kondo localization. In the other extreme where $H_{pair}$ dominates, conduction electrons pair up with each other, resulting in superconductivity. In the present case for LSCO with $x \sim 0.15$, $H_K$ and $H_{pair}$ are both important and in competition, and thereby there is frustration for a conduction electron to pair up either with a local spin or with another conduction electron. Epitaxial strain seems to have a significant influence on this delicate balance. We assume that epitaxial strain dominantly affects the Kondo interaction $H_K$, while leaving $H_{pair}$ nearly unchanged. The Kondo interaction $H_K$ can be expressed as

$$H_K = \sum_i J_K s \cdot S_i$$

Here $s$ represents the spin of a conduction electron (hole of predominantly $O_{2p}$ character), $S_i$ local spin ($Cu^{2+}$ 3d), and $J_K$ the exchange coupling constant between them ($J_K$ is negative for the Kondo effect). How does epitaxial strain regulate $H_K$? Epitaxial strain changes only the lattice parameters ($a_0$ and $c_0$) of LSCO, leaving the doping level constant. The explicit dependence of $H_K$ on these lattice parameters requires microscopic calculations. Our experimental results indicate that shorter $a_0$ and longer $c_0$ reduces either the Kondo coupling constant $J_K$ or the number and/or the magnitude of local spins [14].

Based on the above scenario, we propose a new electronic phase diagram for $La_{2-x}Sr_xCuO_4$ (Fig. 3). In Fig. 3, the thin line represents a *virtual $T_c$ with no Kondo interaction*, the bold line a real $T_c$, and the broken line a Kondo crossover temperature ($T_K$). The gradation of the shadow schematically represents the development of the Kondo effect. Without the Kondo interaction, the virtual $T_c$ ($T^*_c$) would keep increasing with decreasing $x$. The Kondo interaction, however, sets in below a certain $x$ value and becomes stronger for lower $x$. This suppresses the virtual $T_c$ to the real $T_c$, with more significant reduction in $T_c$ for lower $x$, and eventually leads to the disappearance of superconductivity.  As regards the epitaxial strain effect, in-plane compressive strain weakens the Kondo interaction and thereby results in less $T_c$ suppression.   In-plane expansive strain will do in an opposite way.

With overdoping, the Cu-$3d$ orbit becomes more itinerant, resulting in reduction or eventually loss of spin moments.   As a result, the Kondo interaction between a $Cu^{2+}$ local spin and a conduction electron essentially diminishes.   Nevertheless, against the above scenario, $T_c$ decreases by overdoping.   We speculate that the reduction of $T_c$ by overdoping is caused by the weakening of the pairing interaction ($H_{pair}$), although we have no microscopic explanation for this trend at present.

Further, we believe that the pseudogap [15, 16] in the normal state corresponds to the dissociation energy of Kondo singlet bound states.  This implies that the superconducting gap and the normal-state pseudogap are of different origin, but their magnitudes can be comparable at $x_{optimum}$, since both effects are competing.   Finally, from this point of view, we propose the new terminology "Kondo insulator" [17-21] (as opposed to Mott insulator) for underdoped nonsuperconducting $(La,Sr)_2CuO_4$.

In summary, our high-field magnetotransport on "high-$T_c$" and "low-$T_c$" LSCO films with identical chemical composition but with different epitaxial strain demonstrates that superconductivity and the Kondo effect are competing in this compound.  Based on these results, we proposed a new scenario for the electronic phase diagram of LSCO: Kondo interaction suppresses superconducting $T_c$ in the underdoped regime.   Removing or weakening Kondo interaction can raise superconducting transition temperature.

We are obliged to Takashi Hayashi, Akio Tsukada, Dr. Kazuhito Uchida, Dr. Yasuhiro Matsuda, and Prof. Fritz Herlach for technical support and valuable discussions.   This work was partially supported by a Grant-In-Aid for Scientific Research from the Ministry of Education, Science, Sports and Culture, Japan and the New Energy and Industrial Technology Development Organization (NEDO). Correspondence and requests for materials should be addressed to N.M.

with a gap of 30 ~ 200 meV.   As a result, Kondo insulators behave like small-gap semiconductors.

**Table**

Table 1 : $T_c$ and lattice constants of strained La$_{1.85}$Sr$_{0.15}$CuO$_4$ films as compared to the bulk values.

| Sample | $T_c$ | Substrate | $a$-axis ($a_0$) | $c$-axis ($c_0$) |
|---|---|---|---|---|
| "High-$T_c$" film | 44.2 K | LaSrAlO$_4$ | 3.762 Å | 13.29 Å |
| "Low-$T_c$" film | 25.9 K | SrTiO$_3$ | 3.837 Å | 13.18 Å |
| Bulk | 36.7 K | ——— | 3.777 Å | 13.23 Å |

**List of figure captions**

Fig. 1: Comparison of the normal-state resistivity in high magnetic fields ($B$ // $c$-axis) between the "high-$T_c$" and "low-$T_c$" LSCO films.  The "high-$T_c$" film is grown on a LaSrAlO$_4$ substrate, and the "low-$T_c$" film is grown on a SrTiO$_3$ substrate.  The normal-state resistivity of the "high-$T_c$" film is low and metallic down to the lowest temperature, whereas that of the "low-$T_c$" film is high and shows an anomalous semiconducting upturn with a resistivity minimum at around 30 K.

Fig. 2: Resistivity-versus-log $T$ plot of the "low-$T_c$" LSCO film.  The upturn shows a log $T$ dependence, but saturates at lower temperatures.

Fig. 3: New electronic phase diagram proposed for La$_{2-x}$Sr$_x$CuO$_4$.  The thin line represents a *virtual $T_c$ with no Kondo interaction* (conceptual), the bold a real $T_c$, and the broken line a Kondo crossover temperature $T_K$.  The gradation of the shadow schematically represents the development of the Kondo effect.  The edge of the shadowed region is given roughly below the resistivity-minimum temperature. The Kondo interaction suppresses the virtual $T_c$ to the real $T_c$.  In-plane compressive epitaxial strain weakens the Kondo interaction, and thereby results in less $T_c$ suppression.   In-plane expansive strain will do in an opposite way. The symbol e↑=e↓ represents a Cooper pair, and $S$↑=e↓ a Kondo singlet.

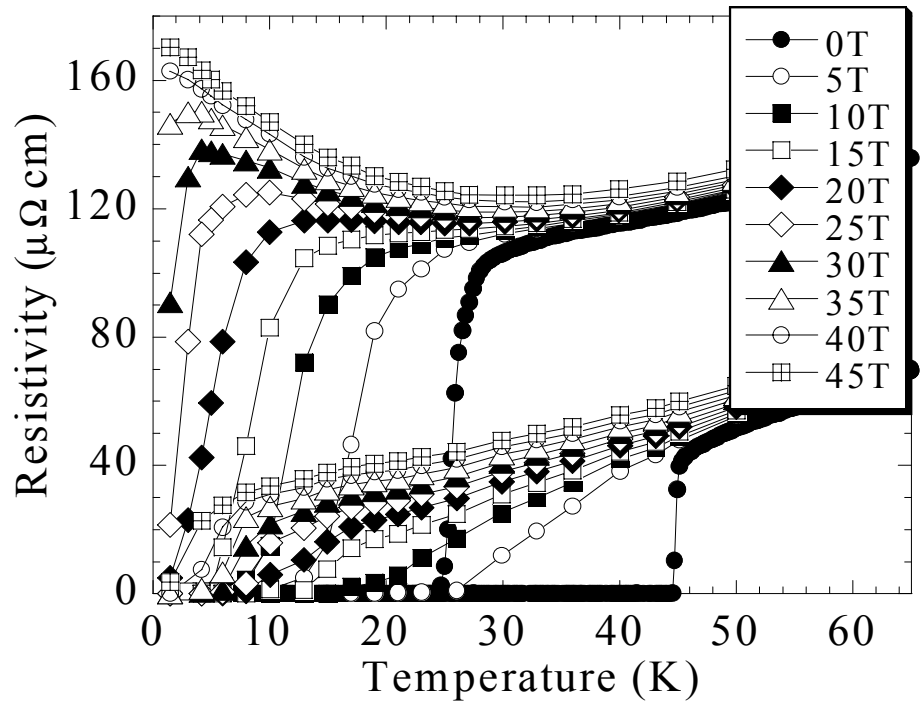

Fig. 1

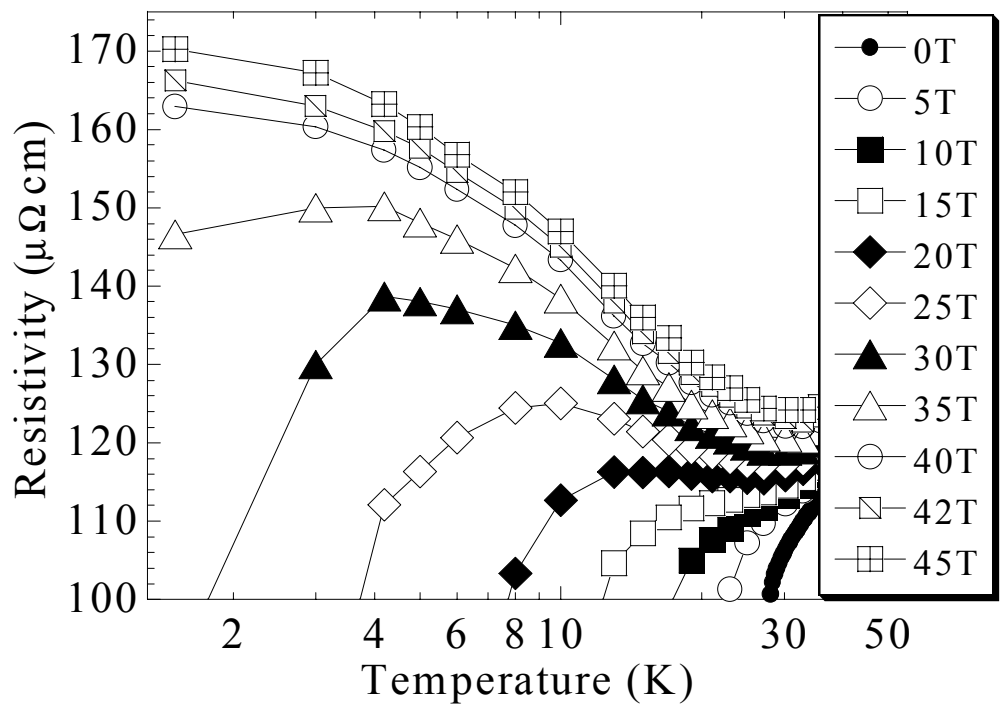

Fig. 2

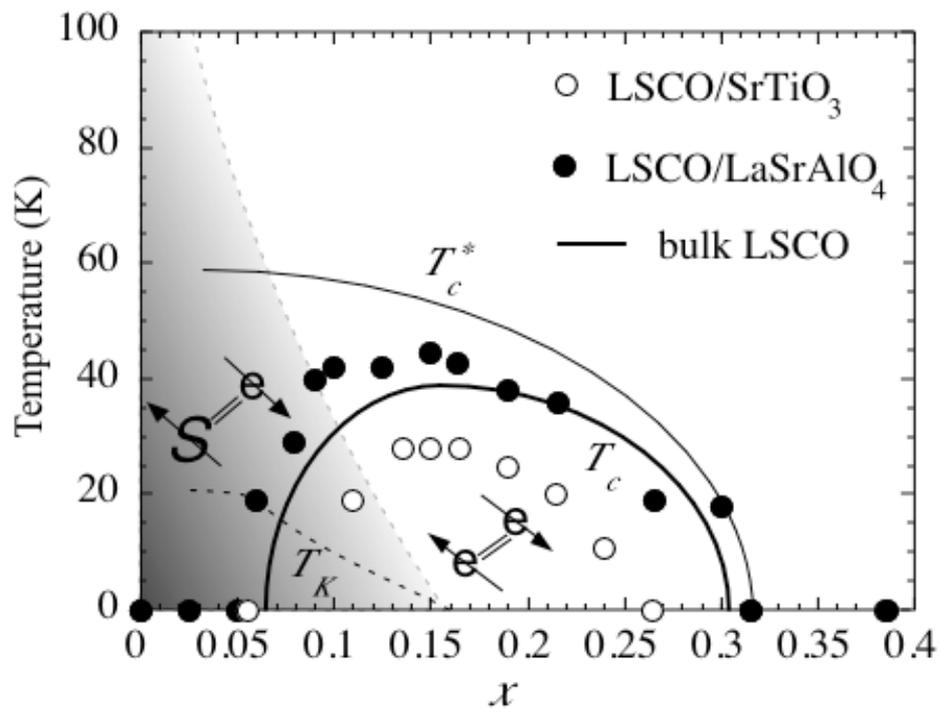

Fig. 3